\documentclass[%
 reprint,
superscriptaddress,
aps,
prb,
 amsmath,amssymb,
]{revtex4-2}

\newcommand{\angstrom}{\textup{\AA}}
\setcitestyle{super}

\usepackage{graphicx, array,multirow }% Include figure files

\usepackage{dcolumn}% Align table columns on decimal point
\usepackage{bm}% bold math
\usepackage{booktabs,xcolor,colortbl}
\usepackage{xparse}
\usepackage[toc,page]{appendix}
\usepackage{caption}
\usepackage{subcaption}
\usepackage{float}
\usepackage{afterpage}

\usepackage{tabularx,booktabs}

\usepackage{palatino}
\usepackage{hyperref}% add hypertext capabilities
\usepackage{todonotes}
\usepackage{color}

\begin{document}

\title{Design of novel organic proton-transfer acid-base (anti-)ferroelectric salts with crystal structure prediction}% Force line breaks with \\

\author{S. Seyedraoufi}
\affiliation{Department of Mechanical Engineering and Technology Management, Norwegian University of
Life Sciences, 1432 Ås, Norway.}

\author{Graeme M. Day}
\affiliation{School of Chemistry, University of Southampton, Southampton SO17 1BJ, United Kingdom.}

\author{Kristian Berland}
\email[E-mail: ]{kristian.berland@nmbu.no}
\affiliation{Department of Mechanical Engineering and Technology Management, Norwegian University of
Life Sciences, 1432 Ås, Norway.}

\date{\today}

\begin{abstract}
Organic molecular ferroelectrics, including organic proton-transfer ferroelectrics and antiferroelectrics, are potentially attractive in organic electronics and have significant chemical tunability. 
Among these, acid-base proton transfer (PT) salts stand out due to their low 
coercive fields and the possibility to tune their properties with different acid-base combinations. 
Using crystal structure prediction, combining small acid and base organic molecular species,
we here predict
three novel acid-base PT ferroelectric salts with higher polarization than existing materials.
We also report two combinations that form antiferroelectric crystal structures.
However, some combinations also result in unfavorable packing or the formation of co-crystal or in one case a divalent salt.
The protonation state is found to be highly linked to the crystal structure,
with cases where salt crystal structures have the same energetic preferability as co-crystals with a different crystalline packing.
\end{abstract}

\maketitle

%\tableofcontents

\section{\label{sec:level1} Introduction}

There is growing interest in 
small-molecule ferroelectric and corresponding antiferroelectric materials.\cite{book,horiuchi_hydrogen-bonded_2020,pan_past_2024}
Traditional ceramic and polymer ferroelectrics find wide usage\cite{appl_modern, Mai2015} in capacitors, sensors, actuators, and memory devices,\cite{appl_modern, Fe_mem} 
due to their piezoelectric and pyroelectric properties, along with their switchable spontaneous polarization ($\bm{P}_{\rm s}$).\cite{applicasss} 
However, small-molecule variants offer
potential advantages over ceramics such as biocompatibility,\cite{li_molecular_2013-1} lightweight, 
low-cost fabrication routes, and integration into flexible substrates.\cite{book, rezbop,alaa, OFE, flex, pan_past_2024} 
Moreover, unlike standard polymer ferroelectrics that suffer from high coercive fields,\cite{pvdf, alaa} they can host fast switching and low-voltage operation.\cite{song, crca,song_bistable_2020}
A particularly attractive aspect of small-molecule ferroelectrics is the immense design flexibility offered through strategies such as functionalization\cite{ai_hf_2021,liu_molecular_2020} and co-crystal engineering.\cite{MOC, smartmatt, bromanilicc}

One type of organic small-molecule ferroelectrics is proton-transfer (PT) ferroelectrics, in which the polarization direction can be reversed through PT between molecules. 
Such systems can be realized both in tautomeric crystals,\cite{cbdc, crca, rezbop, alaa,XAVFEG} amphoteric compounds,\cite{antharnilic} and in organic acid-base salts.\cite{cowqir, mampu, ridfoa, tivdaf, seyed_1}
Compared to the tautomeric crystals,\cite{alaa} the acid-base salts tend to have lower coercive fields, which can be of interest for low-energy switching and sensing.\cite{song_bistable_2020, book}
%, which limits their applications in high-voltage electronics. 
%Attempting to increase $\bm{P}_{\rm s}$, such as by using smaller molecular species, thus,  
%presents an interesting co-crystal engineering challenge. 
Designing novel acid-base PT salts with ferroelectric or antiferroelectric properties
constitutes a significant co-crystal engineering challenge: 
changing the base or acid species---i.e., altering the size, shape, acidity, and functional groups--- would allow for systematic adjustment of functional properties such as $\bm{P}_{\rm s}$,  but can also change the 
preferred synthons, and create packing arrangements that disrupt directional hydrogen bonding\cite{sarma_hydrogen_2014} needed for PT switching. 
Moreover, it can also alter protonation states.\cite{pt_state}

Thus, it is essential to establish how robust the synthons that support PT are to variations of the molecular species.  
The sensitivity to molecular species can be illustrated by the fact that replacing pyrazine in the bromanilic-pyrazine acid-base crystal\cite{tetrameth_no} 
with tetramethyl pyrazine
changes not only the protonation state but also the hydrogen-bonding pattern.\cite{tetrameth} 
So far only 10 acid-base PT ferroelectrics are known,\cite{mampu, cowqir, tivdaf, ridfoa, mampu_1, sponge, bromanilicc} limiting the amount of insight that can be gained from such comparisons. 
These systems all consist of derivatives of haloanilic acid combined with derivatives of phenazine, a tricyclic compound, 
or bipyridine, i.e., two linked pyridine rings,\cite{mampu,cowqir,ridfoa, tivdaf}
with PT between oxygen and nitrogen. 
%\textcolor{blue}{[Would this benefit from a scheme to show the reader the general types of systems? This could be combined (as part b) with a suggested figure in the last paragraph of the introduction, which shows the specific molecules studied here.]}
Among these, the largest reported $\bm{P}_{\rm s} = 5.8~\mathrm{\mu C/cm^2} $ is that of  bromanilic and 2-(3-(pyridin-2-yl)pyrazin-2-yl)pyridinium salt\cite{tivdaf}
far lower than the value of the tautomeric compound croconic acid
$\bm{P}_{\rm s} = 30~\mathrm{\mu C/cm^2}$.\cite{crca}
Moreover, 
none of the 12 compounds uncovered 
in our recent screening study\cite{seyed_1} of the Cambridge Structural Database (CSD) of known crystal structures,\cite{csd, api}
could be classified as acid-base PT salts.
This scarcity of PT acid-base ferroelectrics highlights the need for targeted design of such compounds.  

\begin{figure*}[t!]
\includegraphics[width=0.95\textwidth]{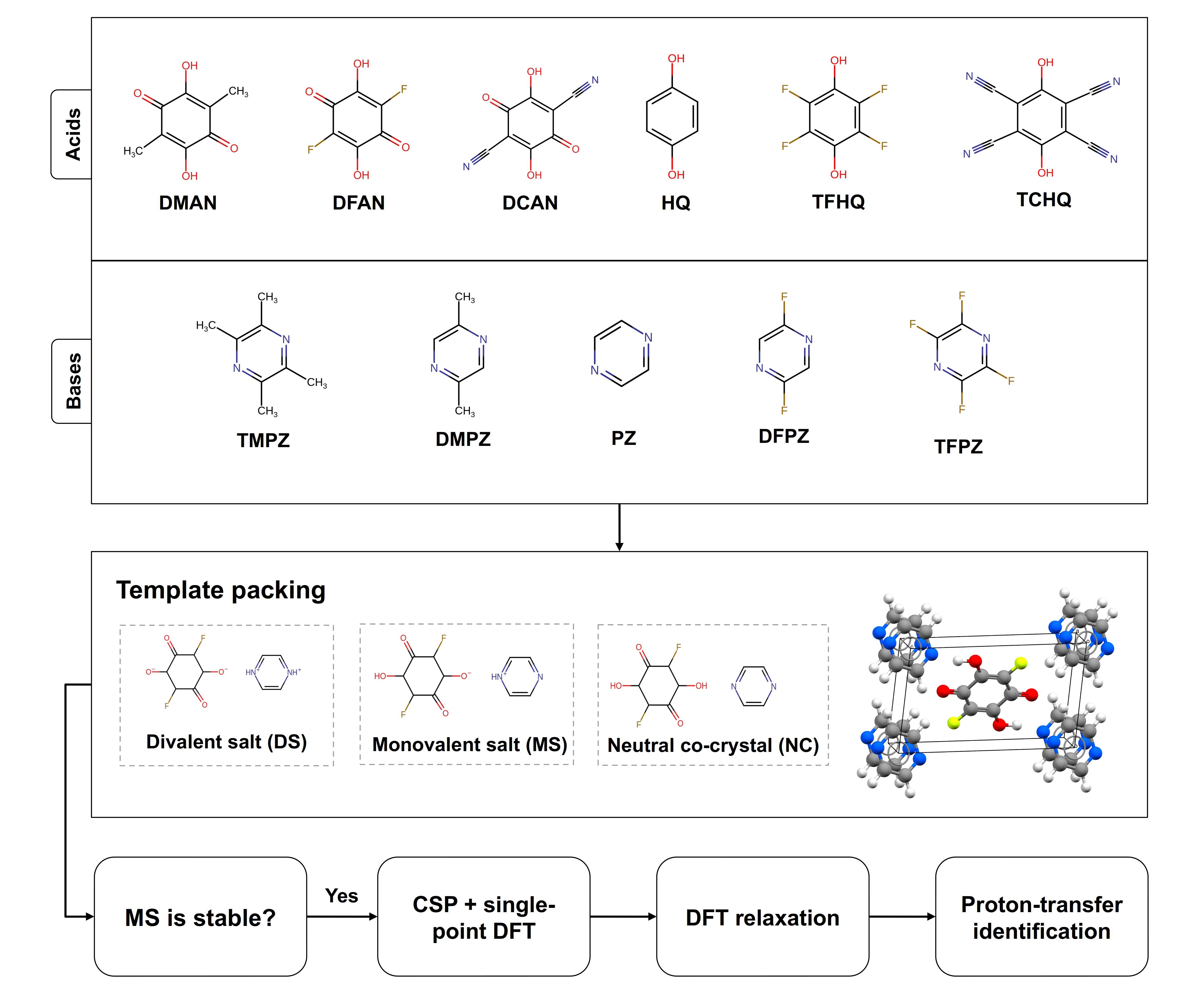}
\caption{\label{fig:workflow} 
The workflow of the acid-base PT salts design, showing the molecules studied (upper panel) and the summary of the computational screening strategy (lower panel).} 
\end{figure*}
%While comparing different experimental studies can provide significant insight, 
%they offer only limited knowledge about competing packing possibilities. 

%Thus, for each new molecular combination, a step to predict the potential solid-state packing is necessary.
Crystal structure prediction (CSP) methods offer a systematic procedure to
identify and assess the relative stability of the numerous possible three-dimensional crystalline arrangements of molecules or atoms from their basic constituents.\cite{blind_test, price_predicting_2014, csp_review}
Each of the local minima on the lattice energy surface is assumed to represent a possible stable crystal structure. CSP %involves exploring the configurational space of crystal structures through 
relies on efficient sampling methods, followed by local energy minimization to identify all relevant local minima. The predicted crystal structures are in turn ranked according to their relative calculated energies,\cite{price_1, Woodley} with the resulting global energy minimum representing the most likely crystal structure to be observed. 
%Somewhat different methods are often used depending on whether the starting point is atoms or molecular species, and here we describe the latter. 
Due to the size of the search space and the number of trial crystal structures that must be assessed, a preliminary CSP landscape is typically first generated with low-cost empirical or semiempirical force fields, limiting the number of systems that must be assessed with DFT.\cite{dmacrys,price_predicting_2014} 
The success of the global optimization approach to CSP using these energy models has recently been assessed in a large-scale study,\cite{1000CSP} demonstrating the reliability of the approach; 74\% of the observed crystal structures for over 1000 molecules are located at, or within 0.02 eV of the global energy minimum from CSP. The reliability of CSP for molecular crystals can be further improved by applying higher-level ab initio methods, usually density functional theory (DFT) including dispersion corrections,\cite{berland_van_2015-1,spin:vdW,pbe-ts, grimme_consistent_2010,otero-de-la-roza_benchmark_2012} 
to a limited number of low-energy structures.\cite{csp_day,csp_review} 

%The outcome of CSP for a molecule or combination of molecules is the structure of the 
%The most energetically favourable crystal structure and the alternative, 
%predicted low energy competing polymorphs, from which properties could be predicted. 
%The results of 
CSP has been successfully applied in a growing number of studies to direct the discovery of new polymorphs, for example of pharmaceutical molecules,\cite{CSPdalcetrapib, Taylor2020} or to guide the functional material design.\cite{pulidoFunctionalMaterialsDiscovery2017, NMOF} 

In this paper, we present a CSP-based design of novel acid-base PT crystals. 
The approach is based on combining different pseudosymmetric single-ring molecular acid and base species that, with the right packing, could potentially support PT between nitrogen and oxygen.
%but limited both bases to different combinations of 
%molecular species that could have equivalent sites on both ends for donating and accepting protons. 
%In this manner, the desired packing would result in infinite PT hydrogen-bonded networks, i.e., they have "connected PT paths".
%Earlier realized acid-base OPTFes, all consist of molecular species where the smallest molecule base is a bicyclic compound or larger. 
We only considered
single-ring molecular species to increase the dipole density and enhance $\bm{P}_{\rm s}$.
We also limited the scope to molecules with methyl, fluoro, cyano, and hydroxyl functional groups, thus totaling 30 combinations, i.e., 6 acids and 5 bases, as shown in the upper panel of Fig.~\ref{fig:workflow}, which describes the workflow of our study.
Based on this screening, we predicted 3 novel ferroelectrics and 2 antiferroelectrics. %\textcolor{blue}{[See comment earlier about adding a figure. This description of the molecules studied in the paper would be clear if there was a figure, where substituents could be shown as X to show where the variations are.]}

%novel acid-base OPTFes using CSP and thus provide insight into how preferable ferroelecric or anti-ferroelectric crystal arrangements are compared to competing crystal structures. 
%In our CSP study, we aimed to increase 
%by 
%However, this replacement changes the effective protonation energy and crystal packing, 
%necessitating a full CSP evaluation.  
%This enhancement is achieved through the substitution of the bulky three-ring base molecule found in conventio/nal acid-base ferroelectrics with smaller, single-ring molecules. 

%The study was based on systematically evaluating 30 novel combinations of acid and base molecules across multiple stages, including ensuring the optimal anti/ferroelectric packing with respect to space group symmetry, protonation state, and PT pathways. 
%Ultimately, we calculated $\bm{P}_{\rm s}$ of the predicted ferroelectric structures, using DFT.

\section{\label{sec:method}  Methods}

\subsection{\label{sec:csp_mine}CSP}
 %Initial geometries of the included species were obtained from crystal structures in CSD\cite{csd, api} with the help of %\textsc{ConQuest}\cite{conquest} and the \textsc{MOLCRYS} Python package\cite{molcris}.  \textcolor{blue}{[I added that this refers to molecular geometries. Is this correct? Otherwise, I don't know what this sentence meant? --> Yes, this is where Mojtaba got the structural motifs from, say rathern than building them from hand. 

The initial stage of crystal structure prediction was performed with rigid molecular geometries, which were optimized using DFT with the 
the second version of the van der Waals density functional (vdW-DF2)\cite{vdw2,berland_van_2015-1, berland_review}, in a box with 15 Å of vacuum padding, using the electrostatic correction scheme of Neugebauer et al.\cite{neugebauer_adsorbate-substrate_1992}.
The configurational space of crystal structures was sampled using the Global Lattice Energy Explorer (\textsc{GLEE}) package,\cite{glee} which employs the Sobol quasi-random method\cite{sobols} to uniformly sample molecular positions, orientations, and unit cell degrees of freedom. Unphysical structures with overlapping molecules were adjusted to relieve clashes, or rejected if clashes could not be relieved. 
Subsequent energy minimization of the trial crystal structures was performed using \textsc{DMACRYS},\cite{dmacrys}
accounting for electrostatic interactions with atom-centered multipoles up to rank 4
and Buckingham-type atomistic potential with dispersion and repulsion terms implemented with interatomic parameters of the \textsc{FIT} potential.\cite{FITT} 
Atomic multipoles were obtained from a distributed multipole analysis\cite{dma} of the charge density obtained with the \textsc{PSI4} software package\cite{psi4} at the B3LYP/6-311G** level.\cite{b3lyp, 611g} 

\begin{figure*}[t!]
 \includegraphics[width=\textwidth]{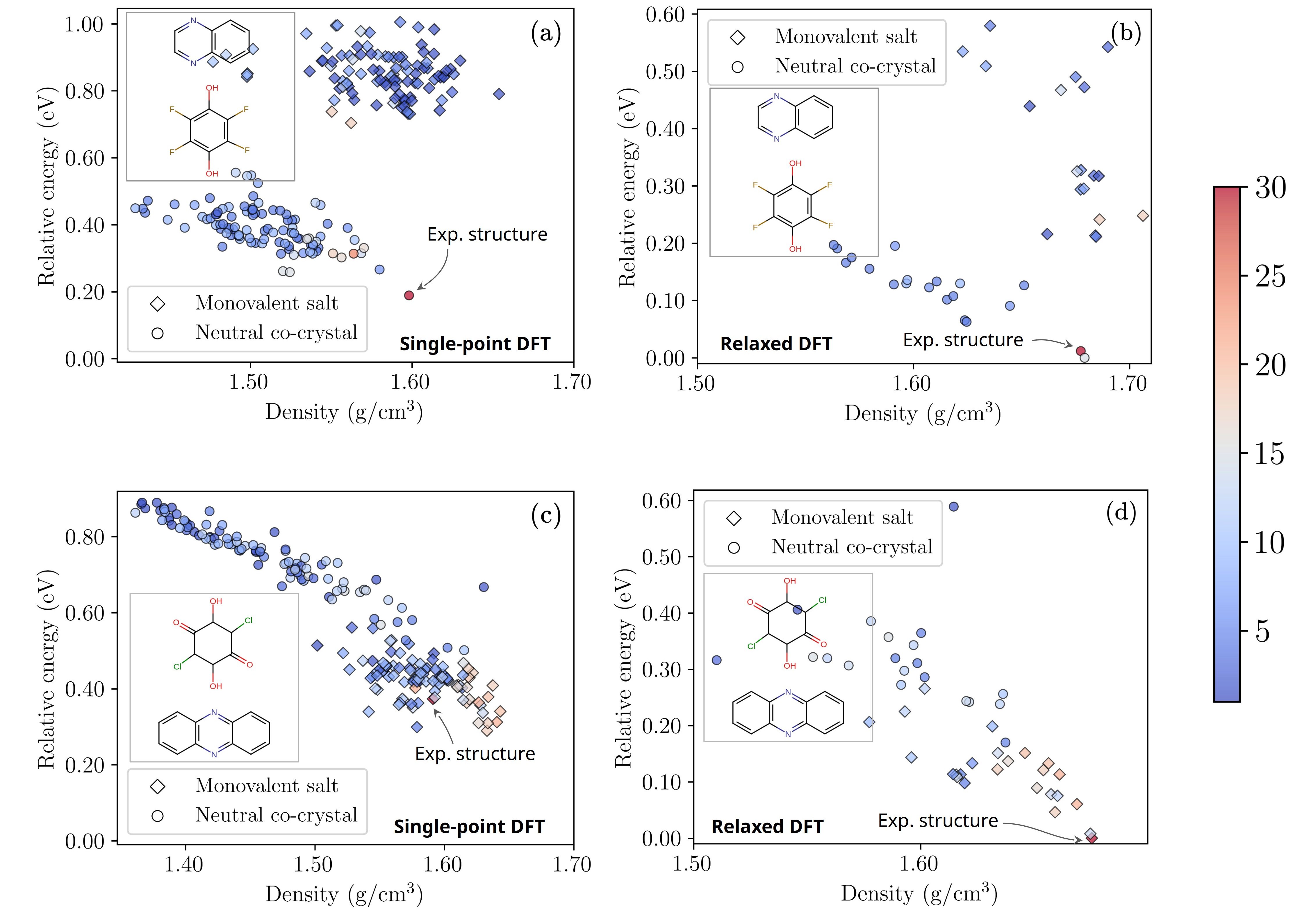}
    \caption{The CSP landscapes for 2,3,5,6-tetrafluorobenzene-1,4-diol quinoxaline (upper panel) and phenazinium chloranilate (lower panel) are plotted using single-point DFT (left) and relaxed DFT (right) energies. The colormap indicates the crystal structure similarity (number of matching molecules) to the experimental using \textsc{COMPACK}.}
        \label{fig:csp_valid}
\end{figure*}

\begin{figure}[t!]
 \includegraphics[width=\columnwidth]{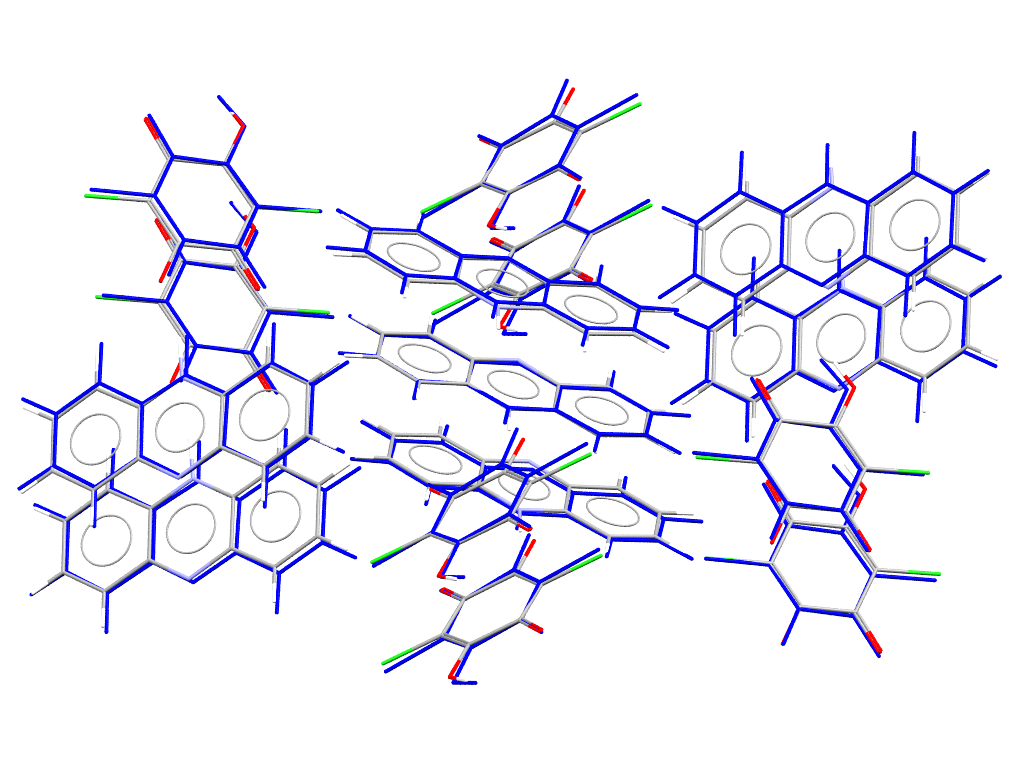}
    \caption{Overlay of the CSP global energy minimum for phenazinium chloranilate (shown in blue) with the experimentally determined crystal structure, CSD reference code MAMPUM03 (shown with atoms coloured by element).}
        \label{fig:MAMPUMoverlay}
\end{figure}

The crystal structure search was performed over the ten most commonly observed space groups for co-crystals, P1, P$\overline{1}$, P2$_1$/c, C2/c, P2$_1$, P2$_1$2$_1$2$_1$, Pbca, Pna2$_1$, C2, and Cc, as well as the common space groups observed for PT ferroelectric crystals, P2/c, Pca2$_1$, Pnn2, Pccn, Fdd2, Iba2, P2, P3, P6$_1$, and I41/a.\cite{alaa, rezbop, cbdc, crca, cowqir, mampu, tivdaf, ridfoa}  
The criteria for terminating the quasi-random generation of crystal structures were either reaching 20,000 successfully energy-minimized structures per space group or failing to generate a new structure after a minimum of 100,000 attempts in that space group. %\textcolor{blue}{[This is unusual. We usually set a target number of structures in a space group and generate trials until we get the required number of successfully energy minimised structures. Was this 20,000 crystal structures in each space group (this is what I expect), and what is the 100,000 attempts?]}
Duplicate structures were removed from the final set of structures by evaluating the similarity of their simulated powder X-ray diffraction patterns generated with \textsc{PLATON}.\cite{platon} 
%This assessment employed the dynamic time-warping method for comparison.
%A second stage of duplicate removal was performed using the \textsc{COMPACK} algorithm,\cite{compack, csd}
%using a cluster of 30 molecules (15 formula units). 
%\textcolor{blue}{[Check that this was performed. --> No, this was not actually performed. I remove it.  ]}
Comparison to experimentally determined crystal structures was also performed using COMPACK, counting the degree of similarity by the number of overlapping molecules
within a 30~\% distance and 30° angle tolerance threshold. 

\subsection{\label{sec:dft_method}Density functional theory calculations}
DFT calculations were performed with the projector augmented wave (PAW) approach, as implemented in the \textsc{VASP} software package.\cite{vasp, vasp1, vasp2} 
A plane-wave energy cutoff of 530~eV was used, which converged the energy difference per acid-base pair between two competing structures in the CSP to within 5~meV. 
As for the single molecules, vdW-DF2 was chosen 
since it could accurately predict the lattice parameters of earlier reported organic PT ferroelectric crystals\cite{seyed_1, chloranilic} and different types of molecular crystals in more general.\cite{chakraborty_next-generation_2020,elin}  
The Brillouin zone was sampled with a $\Gamma$-centered Monkhorst-Pack grid with a minimum k-spacing of ($1/25) \, \angstrom^{-1}$. 
In the electronic optimization, self-consistency was reached when energy differences reached values below $10^{-8}$~eV. 
Ionic relaxations were performed until atomic forces fell below 0.01 eV/Å. 
Finally, the $\bm{P}_{\rm s}$ values were computed using the Berry-phase method,\cite{berry_1, berry_2, berry_3} 
by making a path to a centrosymmetric system using the procedure in Ref.~\onlinecite{seyed_1} using the \textsc{Molcrys} Python tools.\cite{molcris}

\section{\label{sec:results}  Results}

\subsection{CSP validation \label{sec:csp_valid}}
To validate the ability to predict the experimental crystal structures of acid-base PT salts and associated protonation states, we first performed CSP for two systems with known crystal structures. The first, 2,3,5,6-tetrafluorobenzene-1,4-diol quinoxaline (CSD refcode: QUWZIS), is a hydrogen-bonded co-crystal with the P2$_12_12_1$ space group,
which has the same packing as an acid-base salt, but with a neutral paraelectric co-crystal protonation state. 
The second, phenazinium chloranilate (CSD refcode: MAMPUM03),\cite{mampu}  
is an acid-base PT salt crystallizing in the P2$_1$ space group below 254 K,
\cite{mampu_1}
before transitioning to an analogous paraelectric phase. 
CSP was performed for both the neutral co-crystal and monovalent salt states to assess the ability to identify the correct protonation states.

%The other system, QUWZIS, 
Fig.~\ref{fig:csp_valid} shows the CSP landscapes
with the energy per acid-base combination plotted against the density. 
(a) and (b) show the results for molecular combinations with the experimentally favored QUWZIS, respectively, for single-point unrelaxed and relaxed DFT geometries. 
(c) and (d) show the corresponding for the MAMPUM03 acid-base salt.
%The left panels corresponds to DFT calculations with geometries obtained from the force field calulations, while the right corresponds to fully relaxed DFT calculations.
%In both cases, the energies were normalized to that of the lowest relaxed crystal structure, and 
Comparing the unrelaxed and relaxed landscapes shows that the shift due to relaxation can be quite substantial. 

For QUWZIS, a structure matching the experimental coincided with the global minimum single-point DFT energy, which after relaxation shifted to
the second-ranked, but with almost the same energy as the lowest-ranked. 
Moreover, the lowest-ranked also has a rather high similarity to the experimental, nonpolar, and with identical hydrogen-bonding configurations. 
For MAMPUM03, a good match to the experimental study was also identified and ranked 20th in the single-point DFT landscape, but it dropped to the global minimum after relaxations.  
As for QUWZIS, the lowest competing low-energy structure also has a high similarity to the experimental one. The geometrical agreement between the matching CSP structure and the experimental structures is excellent; for QUWZIS, the 30-molecule overlay has a root-mean-square-deviation (RMSD) in atomic positions of 0.126 \AA{}, which for MAMPUM03 (whose overlay is shown in Fig.~\ref{fig:MAMPUMoverlay}), the value is 0.217 \AA.
% KB: I did not see the relvance of this. 
%In the relaxation of 2,3,5,6-tetrafluorobenzene-1,4-diol quinoxaline and phenazinium chloranilate structures were on average displaced by 0.04 eV on the CSP landscape after relaxation compared to the single-point DFT. 

The fact that CSP correctly predicted the experimental structure after relaxation for both validation systems as the lowest or almost so in the energy rankings after relaxation%, and that the nearest competing structures have similar packing, %is highly encouraging. The results give us 
builds confidence in the ability to correctly predict additional acid-base crystal structures, including their protonation state.
Correctly predicting protonation states at the DFT level 
is itself a highly challenging task due to limitations in current exchange-correlation functionals.\cite{leblanc_pervasive_2018,abramov_is_2024}
We speculate that using non-local vdW-DF correlation, particularly vdW-DF2, can improve agreement with experimental results compared to DFT-D methods, due to their strong performance in our recent PT benchmark study\cite{seyed}
%limiting the predictive power of ours results. 
%but our validation study such stimulate to further investigations into the possible role of  non-local correlation at vdW-DF level in conjuction with shape of 
%However, vdW-DFs are generally overlooked in present benchmark studies of molecular crystals. 
%The good performance of vdW-DF2 is possibly linked to its "agressive" exchange enhancement factor\cite{jenkins_reduced-gradient_2021} and repulsive components of the non-local correlation, which we found resulted in accurate PT barriers for molecular dimers.\cite{seyed}
%which could be linked to its fairly accurate PT barriers for molecular dimers.\cite{seyed} 

\begin{figure}[h]
\includegraphics[width=\columnwidth]{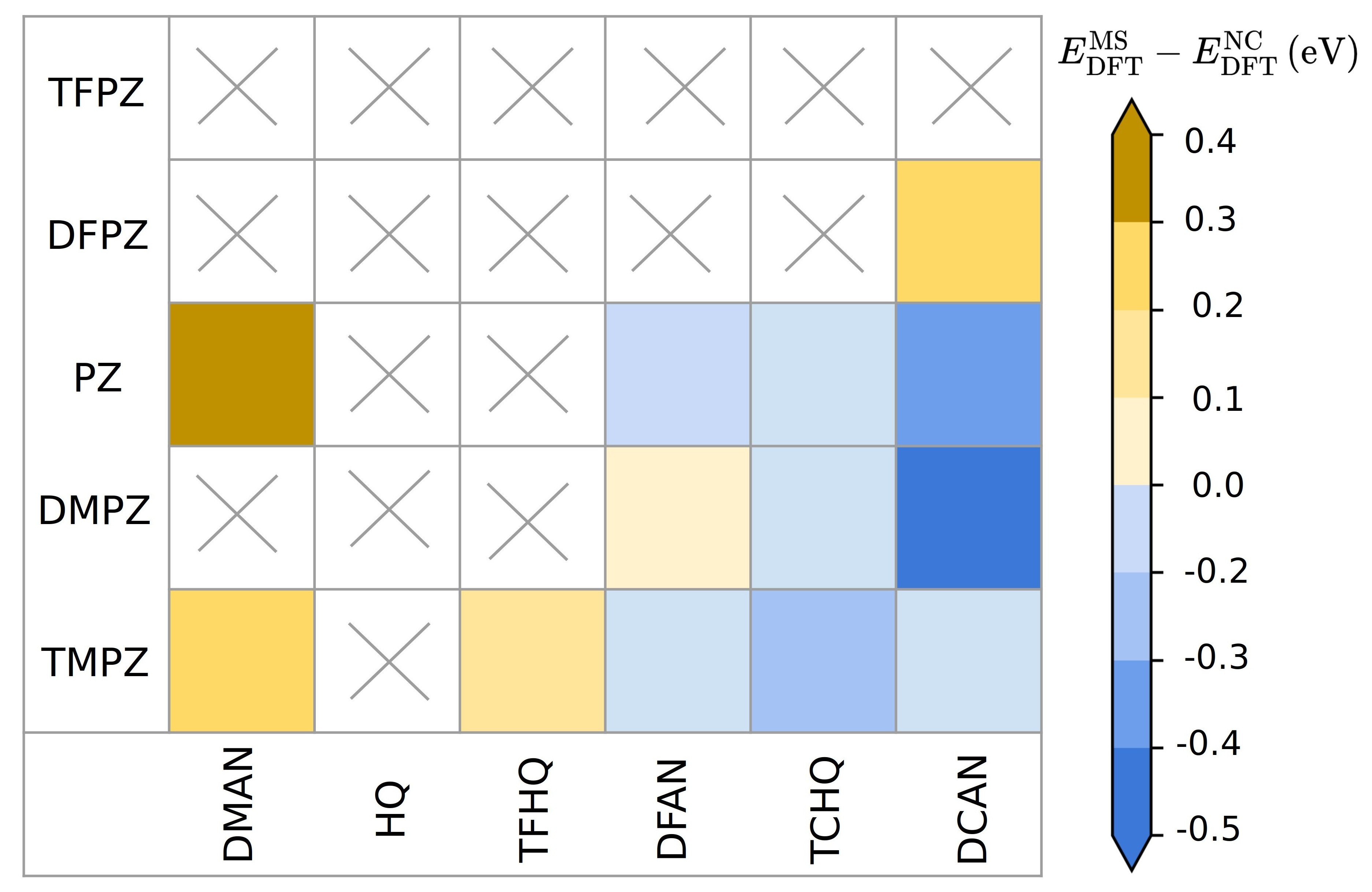}
\includegraphics[width=\columnwidth]{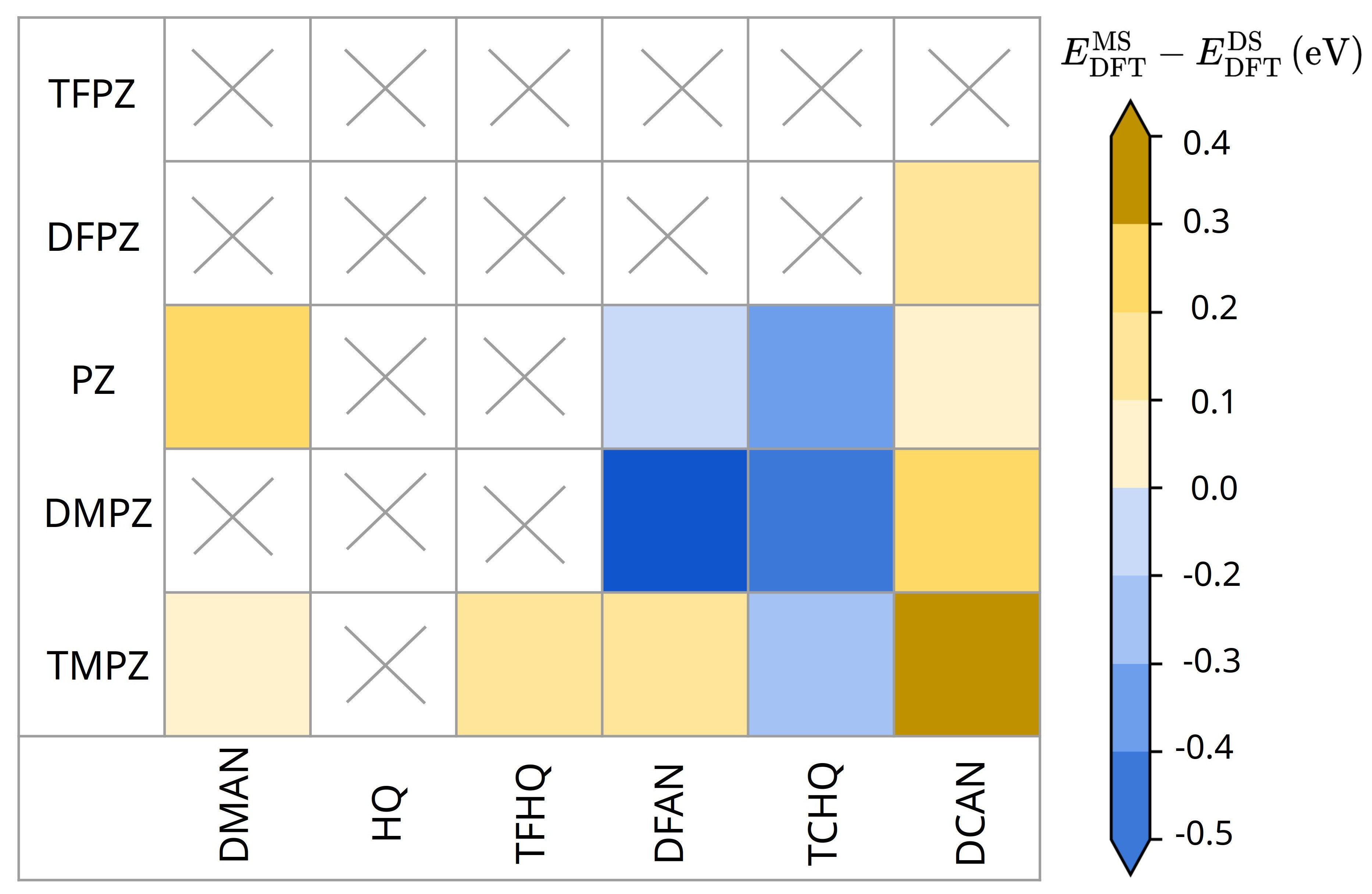}
\caption{\label{fig:grids} 
In the upper panel, the energy difference of MS and NC contour is plotted for different molecular combinations. In the lower panel, the energy difference is plotted between MS and DS.}
\end{figure}

\subsection{\label{sec:csp_design} Co-crystal design}

The design workflow is shown in Fig.~\ref{fig:workflow}. The following acids were considered: \textbf{D}i\textbf{m}ethyl \textbf{an}ilic acid (DMAN), \textbf{d}i\textbf{f}louro \textbf{an}ilic acid (DFAN), \textbf{d}i\textbf{c}yano \textbf{an}ilic acid (DCAN), \textbf{h}ydro\textbf{q}uinone (HQ), \textbf{t}etra\textbf{f}louro \textbf{h}ydro\textbf{q}uinone (TFHQ), and \textbf{t}etra\textbf{c}yano \textbf{h}ydro\textbf{q}uinone (TCHQ);
and the bases:
\textbf{t}etra\textbf{m}ethyl \textbf{p}yra\textbf{z}ine (TMPZ), \textbf{d}i\textbf{m}ethyl \textbf{p}yra\textbf{z}ine (DMPZ), \textbf{p}yra\textbf{z}ine (PZ), \textbf{d}i\textbf{f}louro \textbf{p}yra\textbf{z}ine (DFPZ), and \textbf{t}etra\textbf{f}louro \textbf{p}yra\textbf{z}ine (TFPZ). 

In our study, a key criterion for PT ferroelectricity is the formation of an MS configuration. While PKa and similar criteria can provide some guidance, the protonation state of a molecule can be highly dependent on the environment.
In an attempt to exclude some combinations without having to perform full CSP for all the structures, we first evaluated the protonation states in a relevant crystalline environment.
For this purpose, we considered two different template packings, one of which is depicted in the "template packing" box in Fig.~\ref{fig:workflow}. To accommodate molecules with smaller functional groups such as DFAN, DCAN, PZ, and DFPZ, we adopted a tighter template arrangement with the P1 space group obtained from a similar compound (pyrazine chloranilic acid) deposited in CSD.\cite{Liu2019} 
For systems with base species, a looser packing with the Pc space group from the CSP result of the largest molecules (TCHQ-TMPZ) was employed.

\begin{figure}[t!]
 \includegraphics[width=\columnwidth]{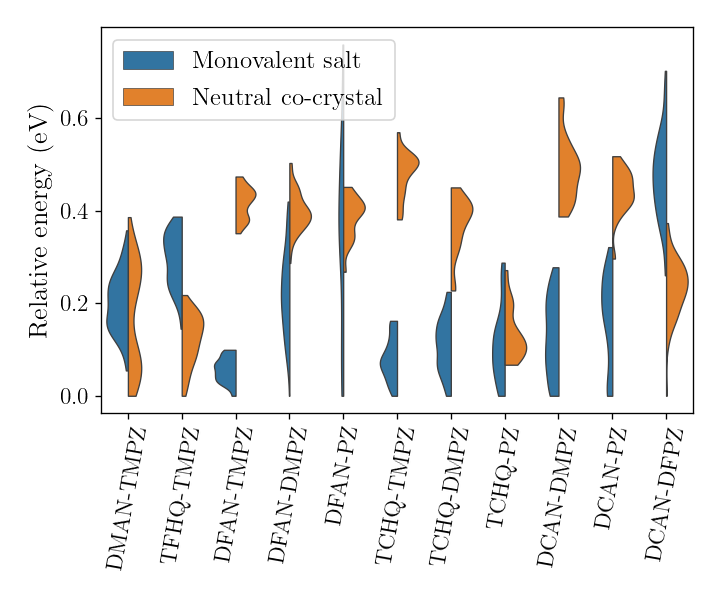}
    \caption{Energy distribution of MS and NC structures generated by CSP using single-point DFT energies for the molecular combinations that successfully passed the template packing step. }
        \label{fig:csp_landscape_spdft}
\end{figure}

Fig.~\ref{fig:grids} shows the resulting heatmap matrices of the energetic favorability of different 
protonation states.
%For the template packings, we evaluated the energy difference between the monovalent salt (MS), 
%the neutral cocrystal (NC), and divalent salt (DS) states for the acid-base %combinations
%after relaxing the atomic positions and cell vectors.  
The upper panel shows the energy difference
between the monovalent salt (MS) and neutral co-crystal (NC), or correspondingly the energy difference between the MS and divalent salt (DS) configurations,  
$E_{\rm DFT}^{\rm MS} - E_{\rm DFT}^{\rm NC (DS)}$,
so that the blue-shaded regions indicate the favorability of the MS state.
For several of the systems, we found the MS state to be dynamically unstable, which are
marked by the $\mathsf{X}$.
With electron-withdrawing substituents, like the cyano group in DCAN and TCHQ, as well as bases with electron-donating groups like methyl in DMPZ and TMPZ, most of the blue shaded boxes in the upper panel appear in the lower right part.
The lower panel shows that the molecular combinations with DCAN exhibit a tendency to donate both of their protons rather than just one, which is consistent with the fact that anilic acid rings tend to be more acidic compared to hydroquinone rings.\cite{hq_pka, cla_pka} 
%As a result of the template packing evaluation,
In the full CSP evaluation, we retained 11 out of the 13 
 combinations exhibiting dynamically stable MS in the template crystal structures.
The exceptions are DMAN-PZ, for which NC is more than $ 0.3$~eV lower per acid-base pair than the MS salt
and DCAN-TMPZ, for which DS is more than $ 0.3$~eV lower than the MS salt. 
%All systems in which DS is preferred were retained, as such structures are less prevalent in experiment.
Initially, we planned to exclude more combinations based on the energetic favorability of the template packing, 
but initial CSP studies 
revealed a substantial impact of the crystal structure on the preferred protonation state, which is also in line with earlier studies.~\cite{pt_state}
However, to limit the computational scope, we generally only performed CSP for the MS and NC protonation states. The exceptions are the cases where a ferroelectric or antiferroelectric packing
are preferred in the global minimum and DS was preferred over MS in the template packing. 
%Thus, some DS might be missed. 
%For DCAN-DMPZ, we also studied the DS state corresponding to the NC optimum,. 
%I looked at the template packing plot, and if DS was more stable in the plot, I

\subsubsection{CSP using single-point DFT energies}

Fig.~\ref{fig:csp_landscape_spdft} compares the energetic distribution of MS and NC crystal structures of the 100 lowest energy structures obtained from CSP with DMACRYS followed by a single-point DFT energy evaluation for the 11 considered combinations. 
For TFHQ-TMPZ and DCAN-DFPZ, most of the low-energy structures
were NCs, with all MS structures having an energy of at least 0.2 eV above the optimal 
NC structure. Based on the validation study, in which both the single-point and relaxed DFT calculations 
correctly identified the protonation state, we deemed it unlikely that MS structures would become energetically favored with a DFT relaxation for these systems, and they were not examined further.
NC was also preferred for DMAN-TMPZ, but in this case by such a small margin that it was retained for full relaxations. 
For all the others, MS was preferred, and they were retained. %Note for the case of DFAN-PZ, there are rater few 

\begin{figure*}[t!]
 \includegraphics[width=0.9\textwidth]{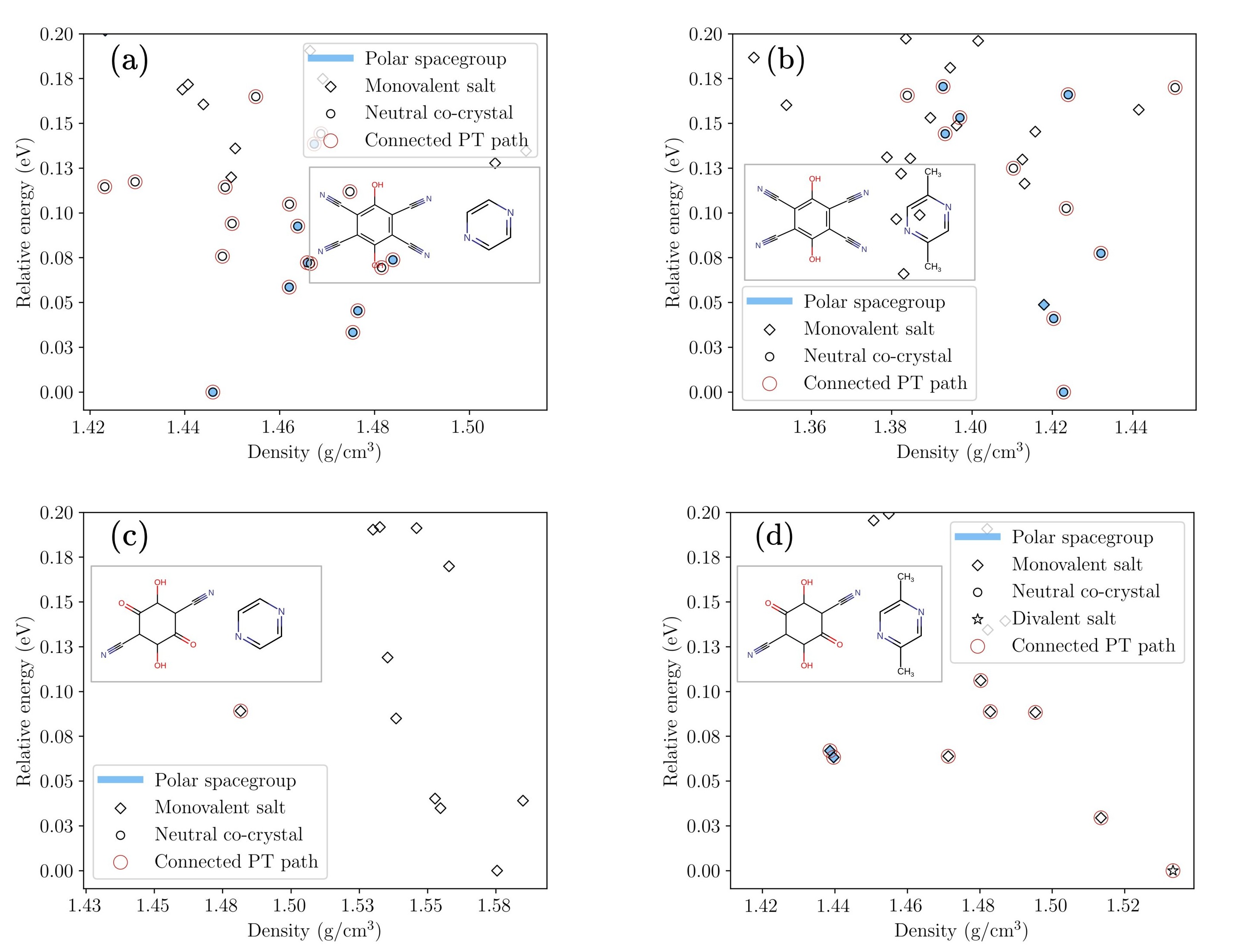}
 
    \caption{The CSP landscapes for (a) TCHQ-PZ, (b) TCHQ-DMPZ, (c) DCAN-PZ, and (d) DCAN-DMPZ molecular combinations are plotted using relaxed structures DFT energies. }
        \label{fig:csp_relax_no}
\end{figure*}

%Thus, we excluded them from further investigations.
%For five of the molecular combinations, DFAN-TMPZ, DFAN-DMPZ, TCHQ-TMPZ, DCAN-DMPZ, and DCAN-PZ, the MS structures were favored over the NC,
%while for DFAN-PZ, DMAN-TMPZ, TCHQ-DMPZ, and TCHQ-PZ, the MS was favored, but by such a small margin that NC were retained in the full relaxation.  
%, as exxcept strognly sof or  

Comparing the favored protonation state in the template packings in the upper panel of Fig.~\ref{fig:grids} with the energy distributions in Fig.~\ref{fig:csp_landscape_spdft}
%we find some notable reorderings highlighting the importance of taking the crystal packing into account. 
shows that the trends are generally maintained. 
However, for DFAN-DMPZ, an MS became preferred, which was also the one with the least energy difference between protonation states in the template packings. 
There is a significant overlap in the distribution of NC and MS structures in some cases, such as for DMAN-TMPZ, further highlighting the link between protonation state and crystal structure. 
%Note that spontaneous changes of protonation states in the final rankings were checked visually for all the low-energy structure after relaxation in the final stage.  

\subsubsection{CSP using relaxed DFT energies}
\begin{figure*}[t!]
 \includegraphics[width=\textwidth]{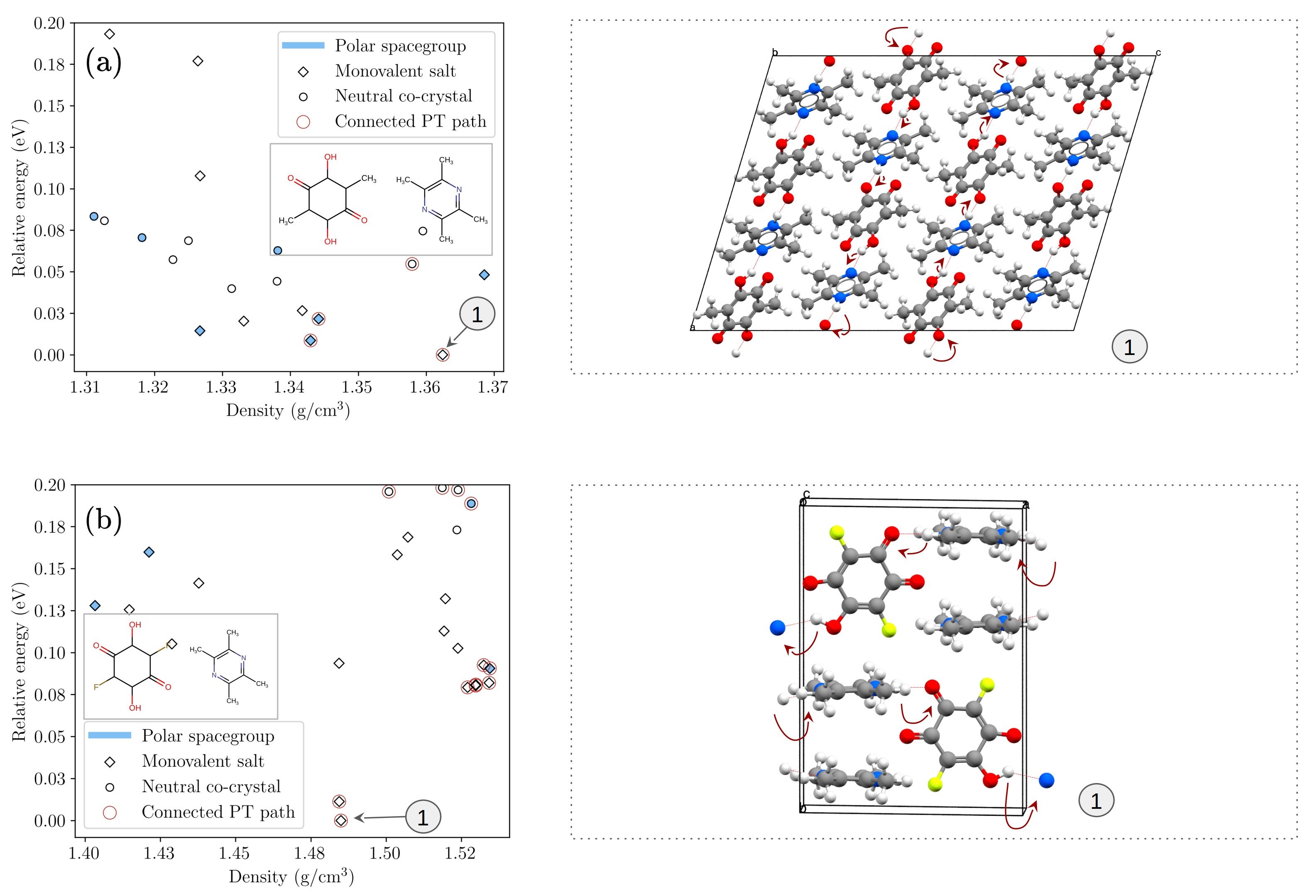}
 
    \caption{The CSP landscapes for (a) DMAN-TMPZ and (b) DFAN-TMPZ molecular combinations are plotted using relaxed structures DFT energies. }
        \label{fig:anti_ferro}
\end{figure*}
For four combinations of molecular species, we found 
global minima after DFT relaxations that were neither PT ferroelectric nor PT antiferroelectrics, as shown in Fig.~\ref{fig:csp_relax_no}.
For both TCHQ-DMPZ (a) and TCHQ-PZ (b), a polar NC crystal structure was preferred 
with connected PT transfer paths. 
Such a structure can, in principle, be ferroelectric,
but with a switching mechanism that is not PT. 
MS salts were also predicted, but none of them exhibited PT paths. 
In the lowest energy MS structures, the negatively charged side of TCHQ forms bonds with the positively charged side of the PZ or DMPZ species, while the neutral hydroxyl ($\rm -OH$) group of TCHQ forms a bond with the cyano group of another TCHQ molecule.
For DCAN-PZ (c) 
the global minima were MS structures without any PT connected path.
For DCAN-PZ (d), there were no MS structures with connected PT paths among the relaxed structures, with more than 0.10 eV energy gap to the closest MS structure with a connected PT path.
For this system, we did not perform any CSP study for DS; 
however, since the template packing clearly favored such a configuration,
we did relax the DS state of this molecular configuration and found it to be more stable than the MS state. As a result, we added the energy of the DS form of the structure in the global minimum to the landscape, which now represents the new global minimum. Other DS crystal structures with even lower energies might exist, but identifying these were not the target of our study.   
%Comparing (c) and (d), it is evident that replacing pyrazine with metal pyrazine has increased the number of MS with a connected PT path, which is important from a design perspective. 

%Once more, we find that the preference for a NC is contrary to the template packings in Fig.~\ref{fig:grids}, but in agreement with the unrelaxed structures in Fig.~\ref{fig:csp_landscape_spdft}.

%The closest MS, which is ranked 17th on the landscape, does not exhibit any connected PT path and is not packed in a polar space group. 
%omparing (a) and (b), we observe that substituting pyrazine with methyl pyrazine has reduced the energy of MS structures on the energy landscape. However, none of the MS structures exhibit connected PT paths. In contrast, all the low-energy NC structures display connected PT paths, and the substitution has not altered this hydrogen bond pattern.

 Fig.~\ref{fig:anti_ferro} shows the CSP landscapes of DMAN-TMPZ (a) and DFAN-TMPZ (b) and the corresponding global minimum crystal structures.
Both have MS global minima with connected PT paths, but nonpolar space groups.
These conditions support PT antiferroelectricity. 
For DMAN-TMPZ, the antiferroelectric global minimum structure has a very small energy gap from %the MS structure with a polar space group and connected PT path, i.e. 
a likely ferroelectric structure, 
while DFAN-TMPZ has a different antiferroelectric crystal structure as the second-ranked structure.
For both these two structures, we performed CSP studies of the DS configurations at the single-point DFT level, since they were preferred in the template packings. 
However, we found the lowest energetic ones to be 0.7 eV and 1.7 eV above the preferred MS structures for DMAN-TMPZ and DFAN-TMPZ, respectively, and they were therefore not subsequently relaxed. 
  
\begin{figure*}[t!]
 \includegraphics[width=\textwidth]{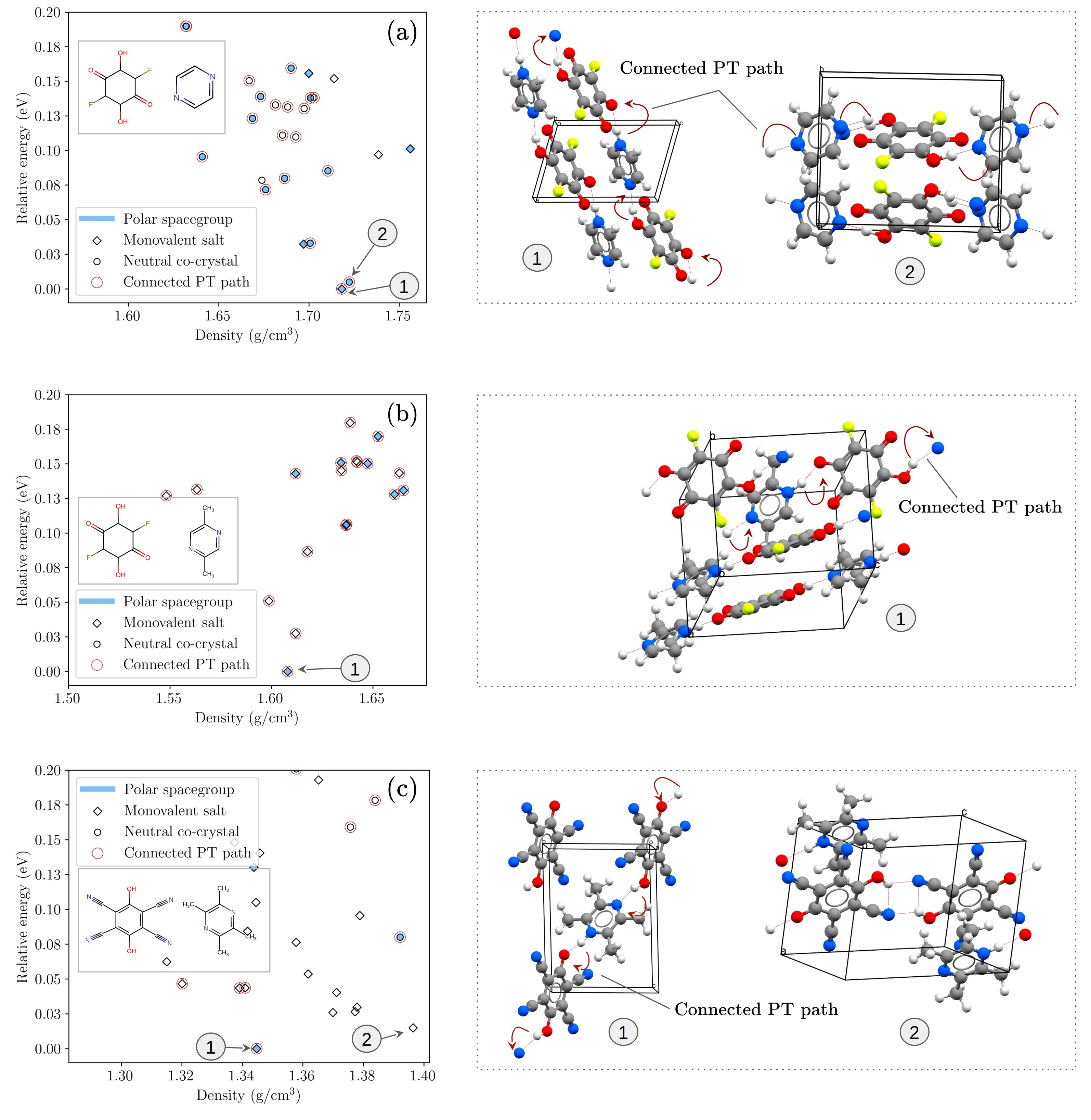}
    \caption{The CSP landscapes for (a) DFAN-PZ and (b) DFAN-DMPZ (c) TCHQ-TMPZ molecular combinations are plotted using relaxed structures DFT energies. }
        \label{fig:ferrosss}
\end{figure*}

Fig.~\ref{fig:ferrosss} shows the CSP landscape and selected associated crystal structures of DFAN-PZ (a), DFAN-DMPZ (b), and TCHQ-TMPZ (c). 
All three have global minima consistent with the conditions of PT ferroelectricity, i.e. MS with connected PT paths. 
%Three molecular combinations exhibit an MS structure with a connected PT path and polar space group in the global minimum. As such  ferroelectric crystals. 
%Panel (a) of , where a ferroelectric candidate is found in the global minima.
For DFAN-PZ (a), most of the low-energy structures have connected PT paths, but many of them are also NC, and the closest structure is just 0.01 eV away from the global minimum.
For DFAN-DMPZ (b) the closest is an antiferroelectric candidate, i.e., an  MS with a centrosymmetric space group P2$_1$/c and connected PT path. All low-energy structures for this combination were either potential ferroelectric or antiferroelectric crystals. 
For TCHQ-TMPZ (c), the low-energy region is mostly occupied by MS structures, but most of them do not have connected PT paths because of the extra synthons that come from the cyano side groups.
Similar to the DFAN-PZ, the energy difference between the global minimum and the second-ranked exhibiting such a structure is merely 0.01 eV.
%Due to the potential for polymorphism, experimental validation is recommended to confirm the stability of the ferroelectric packing structure.
% \begin{table}[t!]
% \caption{\label{tab:table_1} \raggedright Calculated polarizations $|\bm{P}_{\rm cal}|$ ($\rm \mu$C/cm$^2$) are listed for the anti/ferroelectric candidates. }
% \begin{ruledtabular}
% \begin{tabular}{lcc}
% Molecular comb. & $|\bm{P}_{\rm cal}|$ & Space group  \\
% \hline
% \midrule
% \multicolumn{3}{c}{Antiferroelectric candidates} \\
% \midrule
% DMAN-TMPZ & -- & $C2/c$ \\
% DFAN-TMPZ & -- &  $P2_1/c$\\
% \midrule
% \multicolumn{3}{c}{Ferroelectric candidates} \\
% \midrule
% DFAN-PZ & 14.9 & $P1$ \\
% DFAN-DMPZ & 15.0 & $Pc$ \\
% TCHQ-TMPZ & 9.1 & $P1$ \\

% \end{tabular}
% \end{ruledtabular}
% \end{table}
For the ferroelectric systems, we calculated $|\bm{P}_{\rm s}|$ to be 14.9, 15.0, and 9.1~$\rm \mu$C/cm$^2$ for DFAN-PZ, DFAN-DMPZ, and TCHQ-TMPZ, respectively, with the first two having values nearly double those of existing acid-base co-crystals. 
%These values are, on average, three times larger than reported $|\bm{P}_{\rm s}|$ of acid-base OPTFes.    

\section{\label{sec:conclusion} Conclusion and Outlook}
Using DFT and CSP, we evaluated 30 different combinations of acid and base molecular species to identify new acid-base PT salts with enhanced ferroelectric characteristics. In the preliminary stage, 18 combinations were ruled out 
due to the low favorability of the MS state.
%either due to either the lack of a dynamically stable MS protonation state or, in one instance, due to significantly higher NC energy compared to MS energy, as assessed based on the template packingcs. 
After the initial CSP analysis using DFT without relaxations, 3 more candidates were ruled out. Among the remaining 9, two exhibited antiferroelectric packing, and three exhibited ferroelectric packing.

Adding more electronegative groups to the acids and more electron-donating groups to the bases increased the prevalence of MS structures within the low-energy range. Increasing the acidity of acids and the basicity of bases can lead to a DS structure being the most stable, which was the case for one 
of the structures considered in our final CSP ranking. 
Although such considerations of acidity do provide guiding principles, as do considerations of synthon formation, i.e., functional groups in which new synthons emerge can more easily form structures without connected PT paths, 
our study highlights that many types of configurations including different protonation states can arise within the lowest energy structures. This finding highlights the crucial role of the unique environment of each crystal structure and hence also the importance of including CSP in the computational design of small-molecule organic ferroelectrics. CSP using DFT with the vdW-DF2 functional was shown here to accurately predict the structure and protonation state of two validation systems with known structures.

Finally, we hope that the structures we have predicted 
ferroelectric - DFAN-PZ, DFAN-DMPZ, and TCHQ-TMPZ - will be realized and evaluated experimentally to test our predictions, or alternatively one of the systems predicted  antiferroelectric 

\section{Data availability}
A dedicated GitLab page \href{https://gitlab.com/m7582/csp_data}{gitlab.com/m7582/csp$\_$data} describes generated data (with links to different NOMAD uploads), hosts python scripts used in data analysis and plotting, and describes the workflow for generating CSP landscapes. 
The complete NOMAD data set is provided at \href{https://doi.org/10.17172/NOMAD/2024.10.22-1} {10.17172/NOMAD/2024.10.22-1}.

%All the computations and structures for the predicted anti/ferroelectric candidates in this study can be accessed through the NOMAD database at \url{https://dx.doi.org/10.17172/NOMAD/2023.06.01-1}.
%The supporting information is available: Predicted lattice vectors for OPTFe benchmark dataset. All other data is available upon reasonable request. 

\begin{acknowledgments}
KB and SS acknowledge valuable discussions with O. Nilsen, C.H. Görbitz, and M. Balagopalan.
The computations of this work were carried out on UNINETT Sigma2 high-performance computing resources (grant NN9650K and NS8403K). Work by KB and SS is supported by the Research Council of Norway as a part of the Young Research Talent project FOX (302362). GMD thanks the European Research Council for funding, under the European Union’s Horizon 2020 research and innovation program (grant agreement 856405). 
\end{acknowledgments}

\bibliography{ref, from_kb}
\end{document}